\documentclass[aps,prl,twocolumn,showpacs,groupedaddress]{revtex4}  
\usepackage{graphicx}  
\usepackage{dcolumn}   
\usepackage{bm}        
\usepackage{amssymb}   

\begin{document}



\title{Influence of a static magnetic field on the photoluminescence of an ensemble
of\\ Nitrogen-Vacancy color centers in a diamond single-crystal}

\author{Ngoc Diep Lai$^1$}
\author{Dingwei Zheng$^{1,2}$}
\author{Fedor Jelezko$^3$}
\author{Fran\c cois Treussart$^1$}
\author{Jean-Fran\c cois Roch$^1$}
\email{roch@physique.ens-cachan.fr}

\affiliation{$^1$Laboratoire de Photonique Quantique et Mol\'eculaire, UMR CNRS 8537, Ecole Normale Sup\'erieure Cachan, 94235 Cachan cedex, France} 

 \affiliation{$^2$State Key Laboratory of Precision Spectroscopy, East China Normal University, 3663 Zhongshan Road North, Shanghai 200062, China} 

\affiliation{$^3$3. Physikalisches Institut, Universit\"at Stuttgart, Pfaffenwaldring 57, D-70550 Stuttgart, Germany}%

\date{\today}
             

\begin{abstract}
We investigate the electron spin resonance of an ensemble of Nitrogen-Vacancy (NV) color centers in a bulk diamond crystal. The four possible orientations of the NV-center in the lattice lead to different dependences on the magnitude and the orientation of an external static magnetic field. Experimental results obtained with a continuous microwave excitation are in good agreement with simulations. In addition, we observe that the average radiative lifetime of the NV color center is also modified when the external magnetic field is applied. This variation is explained by the mixing between $m_S = 0$ and $m_S = \pm 1$ spin states of the NV-center with different radiative lifetimes, due to magnetic coupling. These results are of interest for a broad range of applications, such as spin-resonance-based magnetometry with a high-density ensemble of NV-centers.
\end{abstract}

\maketitle

Due to its unique features, the Nitrogen-Vacancy (NV) color center in diamond is a promising system for numerous applications. At the individual level, the NV-center is an efficient single-photon emitter.$^1$ Its electron spin properties with exceptional long coherence time$^{2,3}$ can be used to construct quantum gates operating at room temperature$^{4,5,6}$ and to measure magnetic field with nanoscale resolution.$^{7,8}$ At the high-density ensemble level,$^{9}$ NV-centers behave as very sensitive magnetometers on a micrometer scale$^{10,11}$ and are envisioned for building quantum memories where information is coherently in- and out-coupled to spin states.$^{12}$

Due to the $C_{3v}$ symmetry of the NV-center in the diamond crystal, each defect has four possible orientations associated with the [111] axis of the crystal.$^{6}$ Here we discuss the influence of an external static magnetic field on the electron spin resonance of an ensemble of NV-centers. We also measure the magnetic field's influence on the average photoluminescence lifetime.

Figure 1(a) shows the structure of an NV color center in a diamond lattice, fabricated in [110]-orientation. The NV-center consists of a substitutional nitrogen (N) associated to a neighboring vacancy (V). We start from a type-Ib HPHT single-crystal sample. NV-centers are then created from the initially embedded nitrogen impurities, by irradiation with a high-energy electron beam and annealing for 2 hours at 850$^\circ$C. With the applied irradiation dose of $10^{13}$ electron/cm$^2$, a density of about 200 NV-centers/$\mu$m$^3$ can be created. The energy levels of the NV-center are displayed in Fig. 1(b). The NV-center can be optically excited with a laser at a wavelength of 532 nm and emits a broadband luminescence with a zero phonon line at 637 nm. The emission spectrum is measured with an imaging spectrograph, and it shows that the diamond sample mostly contains negatively charged NV$^-$ centers. The ground state of the NV$^-$ center is known to have an electron spin triplet structure with a zero-field splitting of 2.87 GHz between the $m_S = 0$ and the degenerate $m_S = \pm 1$ states.

Optical detection of the NV-center ensemble is realized with a home-built setup. The NV-center is excited with a pulsed laser emitting at 532 nm-wavelength.$^1$ The pulse width is about 800 ps with a 4-MHz repetition rate. The excitation beam is tightly focused on the diamond surface, using a high numerical aperture objective lens (NA = 0.95). The 50-pJ energy per pulse ensures efficient pumping of the NV-centers between ground and excited levels. The photoluminescence of the excited NV-centers is then collected by the same microscope objective and spectrally filtered from the remaining pump light by a long-pass filter with wavelength cut-off at 580 nm. A standard confocal detection system is also used to select the luminescence coming from only a $\sim $1$  \mu$m$^3$ volume in the sample, associated with the excited spot. The photoluminescence signal is finally detected by a silicon avalanche photodiode operating in the single-photon counting regime. 

\begin{figure}[here]
\begin{center}
\includegraphics[width=3.3in]{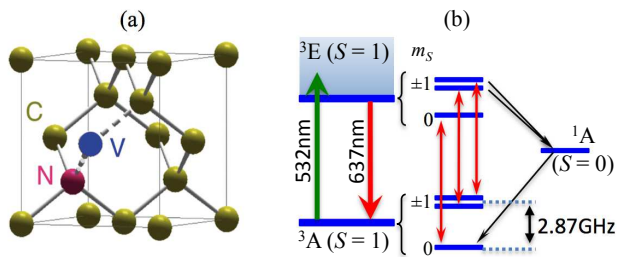}

\caption{(a) Structure of the NV color center in the diamond lattice, consisting of a substitutional nitrogen (N) and a neighboring vacancy (V). (b) Schematic representation of the energy levels for the negatively charged NV$^-$ center. It can be optically excited with a laser at 532-nm wavelength and emits a broadband luminescence with a zero phonon line at 637 nm. The $^3$A ground level is a spin triplet with a zero-field frequency splitting of 2.87 GHz between $m_S = 0$ and $m_S = \pm 1$ states.}
\label{fig:figure1}
\end{center}
\end{figure}
To apply the microwave signal for the electron spin resonance (ESR), the sample is mounted on a circuit board with a microwave strip line for input-output connection. A 20-$\mu$m copper wire placed over the sample is soldered to the strip lines. The wire is positioned to within 20 $\mu$m from the optically addressed NV-centers. The microwave power injected into the strip lines is fixed at 25 dBm for all measurements. The ESR measurement is then realized by sweeping the microwave frequency with a step of 0.3 MHz/channel. The photoluminescence signal is accumulated during a period corresponding to 40 scans of the microwave frequency in order to improve the contrast of the resonance peak. Note that one scan is already enough to identify the magnetic resonance due to the large number of addressed spins within the confocal spot. The external static magnetic field (\textbf{B}) applied to the ensemble of NV-centers can be varied in magnitude and in orientation by moving and rotating a permanent magnet with respect to the sample. For this purpose, the magnet is mounted on translation and rotation stages.

\begin{figure}[here]
\begin{center}
\includegraphics[width=3in]{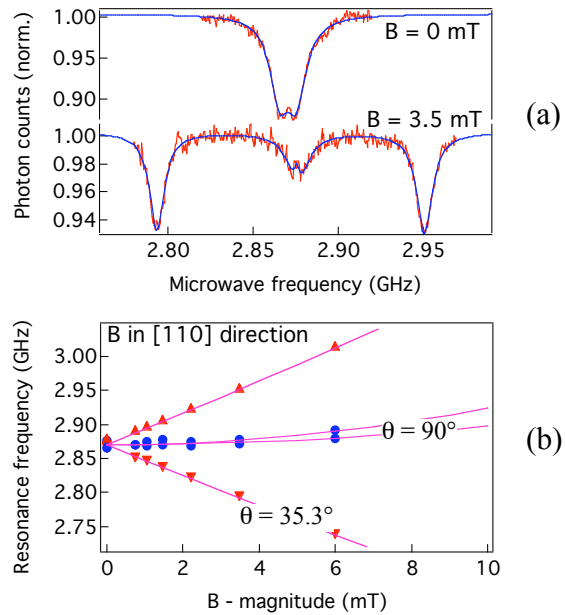}

\caption{(a) Optically detected ESR signal for an ensemble of NV-centers, and the influence of a static magnetic field. (b) Variation of the ESR frequencies associated with the four spin orientations as a function of the magnitude of the applied magnetic field oriented along the [110]-crystal axis. The angles between the NV-spins and the magnetic field are 90$^\circ$ and 35.3$^\circ$, respectively. The small splitting in the zero-field spectrum is due to the fact that the local axial $C_{3v}$ symmetry is broken because of the strain effect, and the degeneracy between $m_S = +1$ and $m_S = -1$ levels is removed.}
\label{fig:figure2}
\end{center}
\end{figure}

Figure 2(a) shows as an example of ESR signal recorded by sweeping the microwave frequency. Without applying the external magnetic field, we observe a drop of luminescence intensity at the microwave frequency of 2.87 GHz due to the induced change in populations of $m_S = 0$ and $m_S = \pm 1$ spin sublevels.$^2$ By applying the static \textbf{B} field, the resonance peak splits by the Zeeman effect. Two resonance peaks are then identified, respectively corresponding to transitions between $m_S = 0 \leftrightarrow -1$, and $m_S = 0 \leftrightarrow +1$ sublevels. By translating the magnet along the [110]-direction, the \textbf{B} field orientation is kept unchanged and the \textbf{B} field magnitude is modified leading to a change in the frequency difference between the two resonance peaks. The frequency of these resonance peaks as a function of \textbf{B}-magnitude is shown in Fig. 2(b). Note that when moving the magnet along the [110]-direction, the magnetic field has the same effect on each of the two groups of NV-centers (shown as 3, 4 and 1, 2 in Fig. 3(a)) among their four possible orientations in the diamond crystal. By fitting$^{13}$ these data, we find out the $\theta$-angle between \textbf{B} and the axis of NV-centers, which are $\theta$(3,4) = 90$^\circ$ and $\theta$(1,2) = 35.3$^\circ$. These values agree with the crystal structure for [110]-oriented sample and the corresponding four possible orientations of the NV-centers indicated on Fig. 3(a).  

\begin{figure}[here]
\begin{center}
\includegraphics[width=3in]{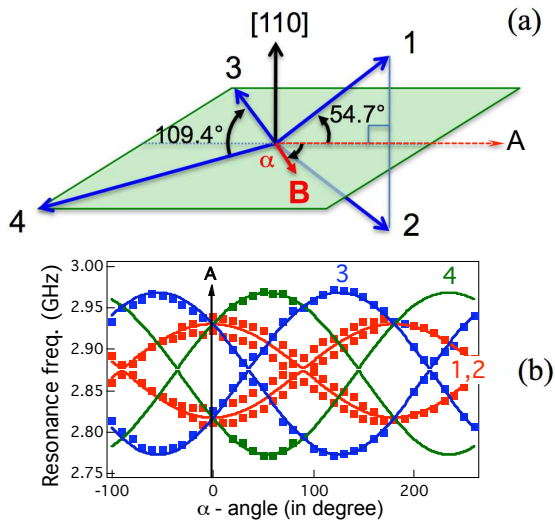}

\caption{(a) Schematic representation of the NV-centers four possible spin orientations in the diamond matrix relative to its [110]-crystal axis. Due to the $C_{3v}$ symmetry, the angle between two possible orientations is 109.4$^\circ$. The laboratory reference axis (A) is chosen to be the symmetrical axis of two possible orientations marked 1 and 2. $\alpha$ is the angle between (A) and the magnetic field when it is oriented to lie in the (3,4)-plane perpendicular to the [110]-crystal axis. (b) ESR frequencies of spins associated with the corresponding orientations 1, 2, 3, and 4, as a function of the rotation angle $\alpha$. Experimental results are shown in dots, and the curves are the simulation results.}
\label{fig:figure3}
\end{center}
\end{figure}

The Zeeman effect depends not only on the magnetic field magnitude but also on its orientation. We therefore carried out a complementary experiment, in which the magnetic field magnitude is kept at a fixed value (B = 3.5 mT), by keeping a constant distance between the magnet and the sample, and in which we move the magnet in such a way that \textbf{B} lies in the plane (3,4) and is rotated around the [110] axis (see Fig. 3(a)). By computing the Zeeman effect for all possible orientations, the ESR frequency is evaluated as a function of the rotation angle $\alpha$. The calculation agrees with the experimental results as shown in Fig. 3(b), except for a small separation observed between the resonance peaks corresponding to the two orientations 1 and 2 of the NV-center. This discrepancy is due to the slight mismatch of the \textbf{B}-field positioning in the plane containing the orientations 3 and 4.

\begin{figure}[here]
\begin{center}
\includegraphics[width=3.3in]{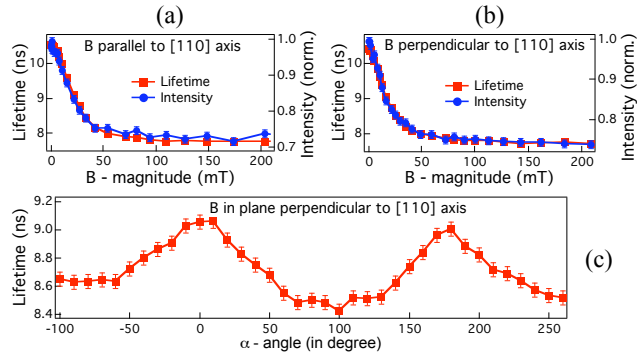}

\caption{Influence of the applied magnetic field on the NV-center average lifetime. The lifetime changes when the external magnetic field is applied, due to the mixing between $m_S = 0$ and $m_S = \pm 1$ states. The magnetic field orientation is parallel (a) or perpendicular (b) to the [110] axis, and the magnetic field magnitude is varied. In (b), the magnetic field direction corresponds to an $\alpha$-angle of $-100^\circ°$. (c) The magnetic field magnitude is fixed at 15 mT and its orientation is rotated in a plane perpendicular to the [110] axis.}
\label{fig:figure4}
\end{center}
\end{figure}

Using the picosecond laser, we can also investigate the influence of the applied magnetic field on the emission rate and on the photoluminescence lifetime. The lifetime is determined by using excitation pulses as starts and photoluminescence photons as stops in the intensity time-correlation measurement setup. Due to different leaking rates toward the spin singlet level  $^1$A,$^2$ the lifetime associated with transition between $m_S = 0$ states is longer than the one associated with transitions between $m_S = \pm 1$ states.$^{14}$ The mixing of these states, induced by the magnetic field, can therefore lead to a modification of the photoluminescence lifetime. By applying the static magnetic field, we indeed observe that the average lifetime for the ensemble of NV-centers is changed, depending on the magnitude and orientation of the \textbf{B}-field. Figures 4(a) and (b) show the dependence of the lifetime on \textbf{B}-magnitude for two particular orientations of the \textbf{B}-field, \textit{i.e.}, parallel and perpendicular to the [110] crystal axis. Since the lifetime associated with $m_S = \pm 1$ states is smaller than the one associated with $m_S = 0$ states,$^{14}$ the lifetime of mixed state due to magnetic coupling has an intermediate value. By recording the photoluminescence signal while changing the \textbf{B}-magnitude, the luminescence intensity is also changed with a clear correlation with the lifetime modification, as shown in Fig. 4(a) and Fig. 4(b). This effect is similarly explained by the mixing of the spin states.$^{6}$ We therefore expect that the measurement of the NV-center lifetime can be applied to either increase the ESR-signal visibility or decrease the ESR measurement time in the case of a weak luminescence signal. Furthermore, the lifetime is sensitive to \textbf{B}-orientation. By keeping \textbf{B}-magnitude constant and changing \textbf{B}-orientation in the plane perpendicular to [110] crystal axis, we indeed observe a correlated modification of the NV-center radiative lifetime, as shown in Fig. 4(c).

However, the variation of the lifetime as a function of the external magnetic cannot be fully simulated, since the number of spins oriented in each of the four possible directions is unknown. Moreover, the excited level of the NV-center is also a triplet state$^{15,16}$ and \textbf{B}-field mixing in this level also contributes to the modification of the average lifetime. 

In conclusion, we have demonstrated the effect of a static magnetic field on the electron spin resonance of an ensemble of NV-centers oriented in the four possible directions in a bulk diamond crystal. The resonance frequency associated with each orientation has a specific dependence on the magnitude and the orientation of the magnetic field. These results have direct application to spin-resonance-based magnetometry with NV centers.$^{7-11}$ Furthermore, the average radiative lifetime of the NV ensemble can be modified by a microwave-free technique, thanks to the mixing of the $m_S = 0$ and $m_S = \pm 1$ spin states by the magnetic field. This effect can be used for the detection of the external magnetic field, without the requirement of a microwave antenna.

This work is supported by the European Commission through EQUIND (FP6 project number IST-034368) and NEDQIT (ERANET Nano-Sci) projects, and by the RTRA ``Triangle de la Physique'' (B-DIAMANT project).

\end{document}